\documentstyle[aps,epsf]{revtex}

\begin{document}
\large{
\title {Time evolution of averages \\
         in dynamical systems driven by noise}

\author{Andrey L. Pankratov  \\
\small{Institute for Physics of Microstructures
of RAS, GSP 105, \\ Nizhny Novgorod, 603600,
RUSSIA. E-mail: alp@ipm.sci-nnov.ru\\}}

\maketitle

\begin{abstract}
This paper presents description of time evolution of
averages of Markov process in wide range of noise intensity.
Exact expression of time scale of average evolution
has been obtained. It has been demonstrated numerically that for
purely noise-induced transitions (transitions over potential barriers)
the time evolution of mean coordinate is a simple exponent with a
good precision even for the case when the potential barrier
height is comparable or smaller than the noise intensity.
Also it has been demonstrated that nonlinear system may be
"linearized" by a strong noise.\\
\\
PACS numbers: 05.40.+j \\
{\it Keywords:} Brownian motion, Markov process,
time evolution of averages.

\end{abstract}
\section{Introduction}
Overdamped Brownian motion in a field of force
(Markov process with nonlinear drift coefficient)
is a model widely used
for description of noise-induced transitions in different
polystable systems. Values that may be observed in experiment are
different averages and knowledge of their time evolution is a
subject of great theoretical and practical importance
in a wide variety of tasks (e.g. tasks of
Josephson electronics \cite{L}, stochastic resonance \cite{st},
ratchet effect \cite{rat1},\cite{rat2}, and so on).

In paper by Nadler and Schulten \cite{Nad} the authors have
demonstrated exponential behavior of observables
for relaxation processes in systems having steady states.
Namely, using the approach of
"generalized moment approximation" the authors of \cite{Nad}
obtained some exact characteristic time scales and for
particular example of a rectangular potential well demonstrated
good coincidence of exponential approximation with numerically
obtained observables.

In the frame of this paper we consider time evolution of
averages of an overdamped Brownian motion.
Extending the approach by Malakhov we have
obtained exact characteristic scales of time evolution of
averages (valid for arbitrary potentials and noise intensity).
Some particular examples of time evolution of mean coordinate
of overdamped Brownian motion in monostable and
bistable potentials have been considered.
It has been demonstrated that time behavior of mean coordinate
is very well approximated by the exponent for purely
noise-induced transitions (transitions over potential barriers)
even in the case of a large noise intensity in comparison with
potential barrier height. Also, an effect of "linearization" of
nonlinear systems by a strong noise has been observed.

It is necessary to mention that in difference from approach used
in \cite{Nad}, our approach allows to describe time evolution of
averages for tasks with arbitrary boundary conditions, e.g. the
case of periodic boundary conditions $\varphi(\pi)=\varphi(-\pi)$
as application to tasks of Josephson electronics will be soon
presented elsewhere. Moreover, as it seems now, the use of the
presented approach for description of ratchet effect may allow
to get the required optimal frequency of flashing \cite{rat1} for
periodic potentials of arbitrary shape.

\section{Main equations and set up of the problem}

    Consider a process of Brownian diffusion in a potential profile
${\mit\Phi(x)}$. Let a coordinate $x(t)$ of the Brownian particle
described by the probability density $W(x,t)$ at initial instant
of time has a fixed value $x(0)=x_0$,
i.e. the initial probability density is the delta function:
\begin{eqnarray}
W(x,0)= \delta(x-x_0).
& &\label{in}
\end{eqnarray}

It is known that the probability density $W(x,t)$ of the Brownian
particle in the overdamped limit (Markov process) satisfies to
the Fokker--Planck equation (FPE):
\begin{eqnarray}
{\partial W(x,t)\over\partial t}=
-{\partial G(x,t)\over\partial x}=
{1\over B}\left\{{\partial\over\partial x}\left[
{d\varphi(x)\over dx}W(x,t)\right]+
{\partial^2W(x,t)\over\partial x^2}\right\}. & &\label{FPE}
\end{eqnarray}
Here $B=\displaystyle{h/{kT}}$, $G(x,t)$ is the probability
current, $h$ is the viscosity (in computer simulations we put
$h=1$), $T$ is the temperature, $k$ is  the Boltzmann constant and
$\varphi(x)=\displaystyle{\mit\Phi(x)/kT}$ is the dimensionless potential
profile. In this paper we will only consider an overdamped Brownian
motion in potentials $\varphi(x)$ such that
${\varphi(\pm\infty)}=+\infty$, i.e. eventually the probability
density will reach a steady-state $W_{st}(x)=Ae^{-\varphi(x)}\ne 0$
($A=1/\int\limits_{-\infty}^{+\infty}e^{-\varphi(x)}dx$).
This leads to the following boundary conditions:
$G(\pm\infty,t)=0$. Besides, the growth of the walls of the
potential $\varphi(x)$ should be fast enough for applicability of
the method described below (for $\varphi(x)\sim x^n$, where
$n\le 1$, the method not always gives correct results).

We will search for the average $m_f(t)$ in the form:
\begin{eqnarray}
m_f(t)=<f(x)>=\int\limits_{-\infty}^{+\infty} f(x)W(x,t) dx.
\label{av}
\end{eqnarray}

Let us mention that the approach we use may be easily extended
for description of time evolution of some kind of conditional
averages when the limits $\pm\infty$ may be substituted by
arbitrary interval $(c,d)$, such that boundary conditions at $c$
and $d$ are arbitrary.

Following this obvious definition (\ref{av}) it is necessary to know the
solution of the FPE - the probability density $W(x,t)$ - for
obtaining the required average. But the nonstationary solution of FPE is
unknown in general case of nonlinear systems. However, as it has been
demonstrated recently in \cite{NP1},\cite{NP2}, the time
evolution of the probability $P(t)=\int\limits_c^d W(x,t) dx$ may be very
well (with only few percent mistake) approximated by the
exponent even when the noise intensity is larger
than a potential barrier height. Besides, it is well-known that
the time evolution of mean coordinate of Brownian motion in
a parabolic potential (linear system) is exactly exponential.
This allows to hope that time evolution of different averages of
Markov process may be in many cases very well described by
exponential approximation in wide range of noise intensity.

Modifying the approach by Malakhov \cite{MP},\cite{M}, it is
possible to define some characteristic scale of time evolution of
the required average (similar way was used in \cite{Nad})
and obtain exact analytic expression for this time scale.
After that, substituting the obtained time scale into the factor
of exponent and comparing the obtained time evolution of the
average with computer simulation results we will test
applicability of this approximation for different types of
potentials in wide range of noise intensity.

\section{Characteristic time scale of evolution of average of Markov
process}

Let us consider attentively the FPE (\ref{FPE}). This is a
continuity equation:
\begin{eqnarray}
{\partial W(x,t)\over\partial t}=
-{\partial G(x,t)\over\partial x}. & &\label{FPEc}
\end{eqnarray}
To obtain necessary average $m_f(t)$ (\ref{av}) let us multiply
equation (\ref{FPEc}) by the function $f(x)$ and integrate it
with respect to $x$ from $-\infty$ to $+\infty$. Then we get:
\begin{eqnarray}
{d m_f(t)\over dt}=
-{\int\limits_{-\infty}^{+\infty} f(x)dG(x,t) }. & &\label{FPEa}
\end{eqnarray}
This is already ordinary differential equation of the first
order ($f(x)$ is known deterministic function), but nobody knows how to
find $G(x,t)$.

Let us define the characteristic scale of time evolution of the average
$m_f(t)$ by similar way as it was done for time scale of the probability
evolution in \cite{MP},\cite{M}:
\begin{eqnarray}
\tau_f(x_0)={\int\limits_{0}^{\infty}\left[m_f(t)-m_f(\infty)\right] dt
\over m_f(0)-m_f(\infty)}.
\label{time}
\end{eqnarray}
Definition (\ref{time}) is general in the sense that it is valid
for any initial condition. But in the frame of this paper we
restrict ourselves by the delta-shaped initial distribution
(\ref{in}) and consider $\tau_f(x_0)$ as function of $x_0$.
For arbitrary initial distribution the required time scale
may be obtained from $\tau_f(x_0)$ by simple averaging over
initial distribution.
If the required function $m_f(t)$ evolves exponentially in time
$m_f(t)\sim e^{-t/\tau}$ then the time scale $\tau$ in the
factor of exponent coincides with $\tau_f(x_0)$ defined by
(\ref{time}).

It is necessary to mention that definition (\ref{time}) gives
correct results only for monotonically evolving functions
$m_f(t)$. Besides,
$m_f(t)$ should fast enough approach its steady-state value
$m_f(\infty)$ for convergence of the integral in (\ref{time}).

The required evolution time of the average $m_f(t)$ may be
obtained via the approach proposed by Malakhov
\cite{MP},\cite{M}. This approach is based on the Laplace
transformation method
$m_f(s)=\int\limits_0^{\infty}m_f(t)e^{-st}dt$.
We will only slightly modify this approach with respect to our
task.

Performing Laplace transform of formula (\ref{time}),
Eq. (\ref{FPEa}) (Laplace transformation of (\ref{FPEa}) gives:
$sm_f(s)-m_f(0)=-\int\limits_{-\infty}^{+\infty}f(x)d\hat{G}(x,s)$)
and combining them, we get:
\begin{eqnarray}
\tau_f(x_0)=\lim_{s\to 0}{sm_f(s)-m_f(\infty)
\over s[m_f(0)-m_f(\infty)]}=
\lim_{s\to 0}{m_f(0)-m_f(\infty)-
\int\limits_{-\infty}^{+\infty}f(x)d\hat{G}(x,s)
\over s[m_f(0)-m_f(\infty)]}.
\label{tl}
\end{eqnarray}
where $\hat{G}(x,s$) is the Laplace
transformation of the probability current $\hat{G}(x,s)=
\int\limits_{0}^{\infty}G(x,t)e^{-st}dt$.

Following the approach by Malakhov, one can
introduce the function $H(x,s)\equiv s\hat{G}(x,s)$,
and expand it in the power series in $s$:
\begin{equation}
H(x,s)\equiv s\hat{G}(x,s)=H_0(x)+sH_1(x)+s^2H_2(x)+ \ldots
\label{a1}
\end{equation}

It is possible to find the differential equations for $H_n(x)$
(see \cite{M}):
\begin{equation}
\begin{array}{ll}
\displaystyle{\frac{dH_1(x)}{{dx}}}=-W_{st}(x)+\delta(x-x_0), &  \\
\displaystyle{\frac{d^2 H_n(x)}{{dx^2}}}+{\frac{d\varphi(x)}{{
dx}}}{\frac{dH_n(x)}{{dx}}}=BH_{n-1}(x),
\quad n=2,3,4,\ldots  \label{a3}
\end{array}
\end{equation}
Here $W_{st}=Ae^{-\varphi(x)}$,
$A=1/\int\limits_{-\infty}^{+\infty}e^{-\varphi(x)}dx$.
Using the natural boundary conditions $G(\pm\infty,t)=0$,
one can obtain from (\ref{a3})
$H_1(x)=-A\int\limits_{-\infty}^{x}e^{-\varphi(v)}dv+1(x-x_0)$ and
\begin{eqnarray}
H_2(x)=B\left\{\int\limits_{-\infty}^{x}e^{-\varphi(v)}
\int\limits_{-\infty}^{v}e^{\varphi(y)}[1(y-x_0)-F(y)]dydv
- \right. \nonumber \\
\left.-F(x)\int\limits_{-\infty}^{+\infty}e^{-\varphi(v)}
\int\limits_{-\infty}^{v}e^{\varphi(y)}[1(y-x_0)-F(y)]dydv
\right\},
\label{H}
\end{eqnarray}
where
\begin{equation}
F(u)=\int\limits_{-\infty}^{u}e^{-\varphi(v)}dv/
\int\limits_{-\infty}^{+\infty}e^{-\varphi(v)}dv.
\label{F}
\end{equation}

Here we restricted ourselves by obtaining of $H_2(x)$ only,
because substituting the set (\ref{a1}) into formula (\ref{tl})
one can get:
\begin{eqnarray}
\tau_f(x_0)=-{\int\limits_{-\infty}^{+\infty}f(x)dH_2(x)
\over [m_f(0)-m_f(\infty)]}.
\label{tlh}
\end{eqnarray}

Thus, substituting $H_2(x)$ (\ref{H}) into formula (\ref{tlh}) one can get
the characteristic scale of time evolution of any average
$m_f(t)$ for arbitrary potential such that
$\varphi({\pm\infty})=\infty$:
\begin{eqnarray}\nonumber
\tau_f(x_0)=\frac{B}{m_f(0)-m_f(\infty)}\times \\
\times \left\{
\int\limits_{-\infty}^{\infty}f(x)e^{-\varphi(x)}
\int\limits_{x_0}^{x}e^{\varphi(u)}F(u)dudx-
A\int\limits_{-\infty}^{\infty}f(x)e^{-\varphi(x)}dx
\int\limits_{-\infty}^{\infty}e^{-\varphi(x)}
\int\limits_{x_0}^{x}e^{\varphi(u)}F(u)dudx+ \right. \label{tf}\\
\nonumber \left.
+A\int\limits_{-\infty}^{\infty}f(x)e^{-\varphi(x)}dx
\int\limits_{x_0}^{\infty}e^{-\varphi(x)}
\int\limits_{x_0}^{x}e^{\varphi(u)}dudx-
\int\limits_{x_0}^{\infty}f(x)e^{-\varphi(x)}
\int\limits_{x_0}^{x}e^{\varphi(u)}dudx
\right\},
\end{eqnarray}
where $F(x)$ is expressed by (\ref{F}) and
$A=1/\int\limits_{-\infty}^{+\infty}e^{-\varphi(x)}dx$.

Once we know the required time scale of the evolution of average,
we can present the required average in the form:
\begin{equation}\label{avt}
m_f(t)=(m_f(0)-m_f(\infty))\exp(-t/\tau_f(x_0))+
m_f(\infty).
\end{equation}
The applicability of this formula for several examples of time
evolution of mean coordinate $m(t)=<x(t)>$ will be checked in the next
section.

\section{Time evolution of mean coordinate of Markov process}

As an example of description presented above, let us consider
time evolution of mean coordinate of Markov process:
\begin{equation}
m(t)=<x(t)>=\int\limits_{-\infty}^{+\infty} xW(x,t) dx.
\end{equation}

The characteristic time scale of evolution of the mean coordinate
in general case may be easily obtained from (\ref{tf}) substituting
$x$ for $f(x)$. But for symmetric potentials $\varphi(x)=\varphi(-x)$
the expression of time scale of mean coordinate evolution may be
significantly simplified ($m(\infty)=0$):
\begin{equation}
\tau_m(x_0)=\frac{B}{x_0}\left\{
\int\limits_{0}^{+\infty}xe^{-\varphi(x)}dx\cdot
\int\limits_{0}^{x_0}e^{\varphi(u)}du+
\int\limits_{0}^{x_0}xe^{-\varphi(x)}
\int\limits_{x_0}^{x}e^{\varphi(u)}dudx
\right\}.
\label{sym}
\end{equation}
If $x_0=0$ then, as not difficult to check, $\tau_m(x_0)=0$.

Let us consider now some examples of symmetric potentials and check
applicability of exponential approximation:
\begin{equation}\label{mt}
m(t)=<x(t)>=x_0\exp(-t/\tau_m(x_0)).
\end{equation}

First we should consider time evolution of mean coordinate in
monostable parabolic potential $\varphi(x)=ax^2/2$ (linear
system), because for this case time evolution of mean is known:
\begin{equation}
m_{par}(t)=x_0\exp(-at/B),
\label{par}
\end{equation}
where $a=a'/kT$ and $B=h/kT$, so $\tau_{par}=B/a$ for linear system
and does not depend on noise intensity and the coordinate of initial
distribution $x_0$. On the other hand, $\tau_m(x_0)$ is expressed
by formula (\ref{sym}). Substituting parabolic potential
$\varphi(x)=ax^2/2$ in formula (\ref{sym}), making simple
substitutions and changing order of integrals, it can be easily
demonstrated that $\tau_m(x_0)=B/a=h/a'=\tau_{par}$, so for
purely parabolic potential the time of mean evolution (\ref{sym})
is independent of both noise intensity and $x_0$ as it must.
This fact proves the correctness of the used approach.

The second considered example is described by the monostable
potential of the fourth order: $\varphi(x)=ax^4/4$. In this
nonlinear case the applicability of exponential approximation
significantly depends on the location of initial distribution
and the noise intensity.
Nevertheless, the exponential approximation of time evolution of
the mean gives qualitatively correct results and may be used as
first estimation in wide range of noise intensity (see Fig. 1,
$a=1$). Moreover, if we will increase noise intensity further,
we will see, that error of our approximation decreases and
for $kT=50$ we get that the
exponential approximation and the results of computer simulation
coincide  (see  Fig.  2, plotted in the logarithmic scale, $a=1$,
$x_0=3$).
From this plot we can conclude that nonlinear system is
"linearized" by strong noise - effect which is qualitatively
obvious, but should be investigated further by the analysis of
variance and higher cumulants.

The third considered example is described by the bistable
potential - the so-called "quartic" potential:
$\varphi(x)=ax^4/4-bx^2/2$. In this case the applicability of
exponential approximation also significantly depends on the coordinate
of initial distribution. If $x_0$ is far from the potential minimum,
then there exists two characteristic time scales: fast dynamic
transition to potential minimum and slow noise-induced escape
over potential barrier. In this case exponential approximation
gives not so adequate description of temporal dynamics of the
mean, however may be used as first estimation. But if $x_0$
coincide with the potential minimum, then the exponential
approximation of the mean coordinate only few percent differs
from results of computer simulation even in the case when noise
intensity is significantly larger than the potential barrier height
(strongly nonequilibrium case) (see Fig. 3, $a=1$, $b=2$,
$x_0=1.414$).
If, however, we consider the case when the initial distribution
$x_0$ is far from the potential minimum, but the noise intensity is
large, we will see again as in previous example that
essential nonlinearity of the potential is suppressed by strong
fluctuations and the evolution of the mean coordinate becomes
exponential (see Fig. 4, plotted in the logarithmic scale,
$a=1$, $b=2$, $x_0=2.5$).

\section{Conclusion}

In the frame of this paper we have considered time evolution of
averages of an overdamped Brownian motion.
Extending the approach by Malakhov we have
obtained exact characteristic scales of time evolution of
averages (valid for arbitrary potentials and noise intensity).
Some particular examples of time evolution of mean coordinate
of overdamped Brownian motion in monostable and
bistable potentials have been analyzed.

It has been demonstrated that for purely
noise-induced transitions (transitions over potential barriers)
the time evolution of mean coordinate is a simple exponent with
a good precision even for the case when the potential barrier
height is comparable or smaller than the noise intensity.

Also,  there  has  been observed the effect of "linearization" of
nonlinear system by a strong noise in the sense that time evolution
of  mean  coordinate  becomes purely exponential with increase of
noise intensity.

\section{Acknowledgments}

Author wishes to thank Prof. A.N.Malakhov for attentive
reading of the manuscript and constructive
comments, and Prof. V.N.Belykh for helpful discussions.

This work has been supported by the Russian
Foundation for Basic Research (Project N~96-02-16772-a,
Project N~96-15-96718 and Project N~97-02-16928), by Ministry
of High Education of Russian Federation (Project N~3877) and
in part by Grant N~98-2-13 from the International Center
for Advanced Studies in Nizhny Novgorod.

\newpage

\begin{figure}[th]
\centerline{
\epsfxsize=12cm
\epsffile{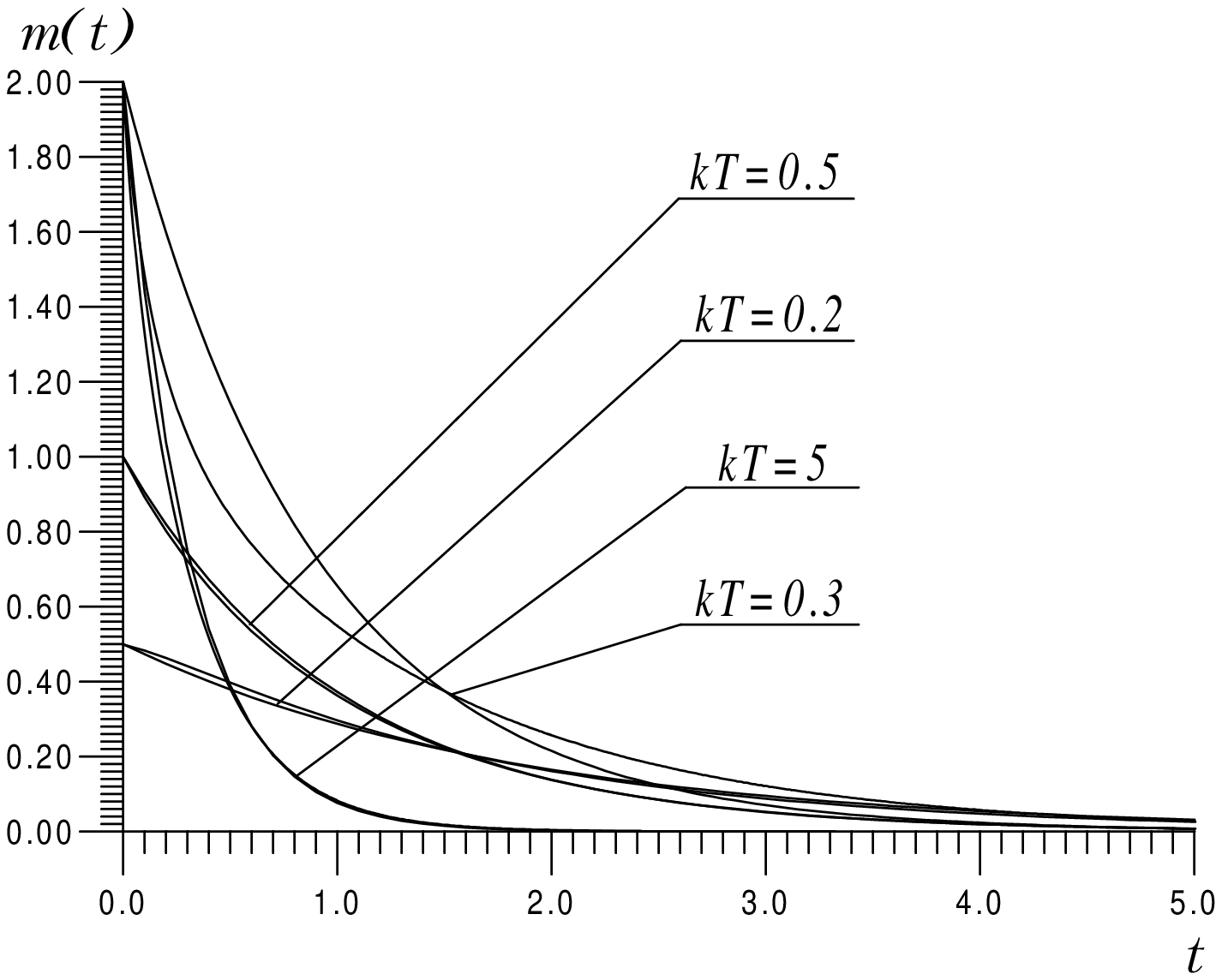}}
\vspace{-55pt}
\caption[b]{Evolution of the mean coordinate in the potential
${\mit\Phi(x)}=x^4/4$
for different values of noise intensity.
}
\end{figure}

\newpage

\begin{figure}[th]
\centerline{
\epsfxsize=12cm
\epsffile{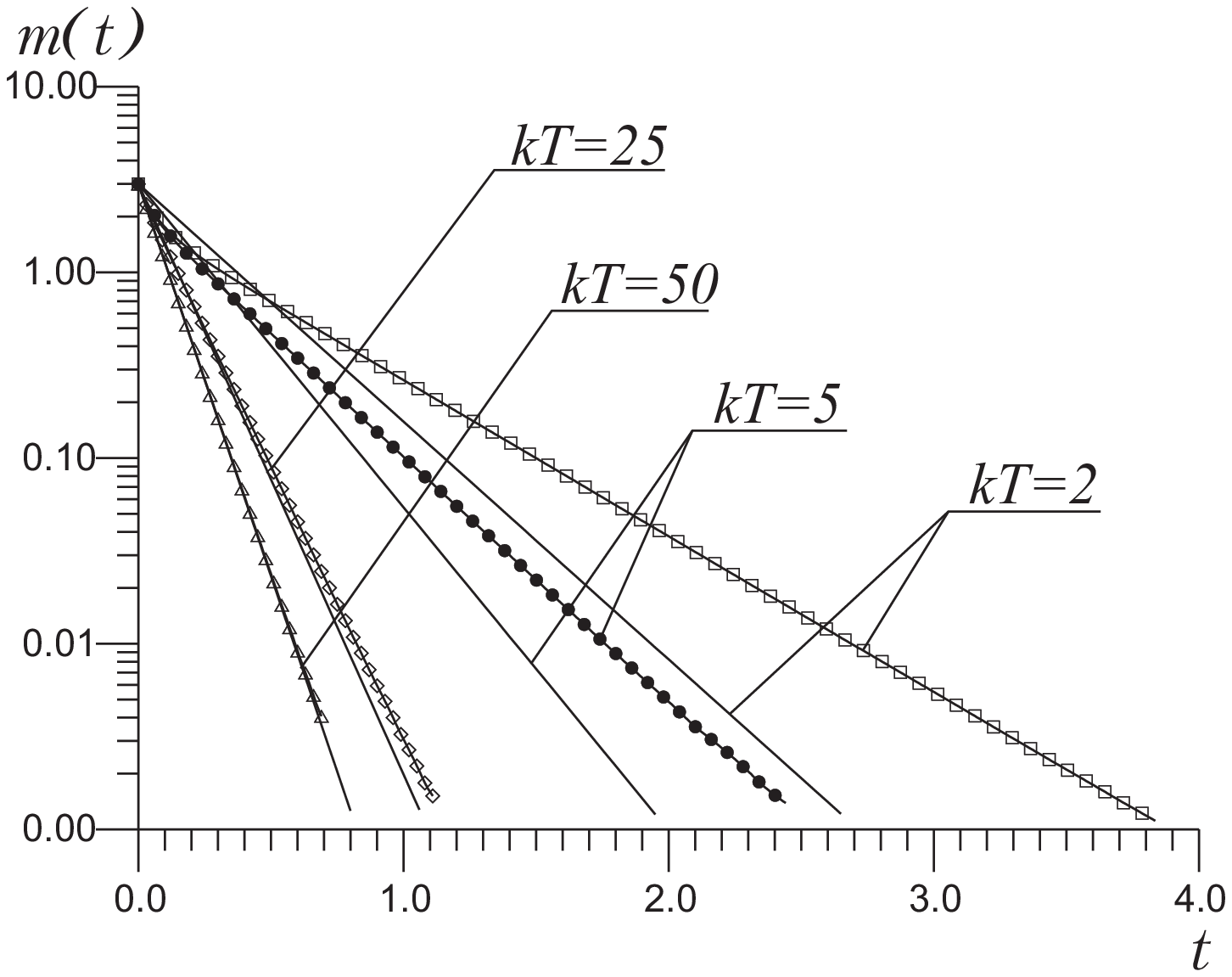}}
\vspace{-55pt}
\caption[b]{Evolution of the mean coordinate in the potential
${\mit\Phi(x)}=x^4/4$
for different values of noise intensity. }
\end{figure}

\newpage

\begin{figure}[th]
\centerline{
\epsfxsize=12cm
\epsffile{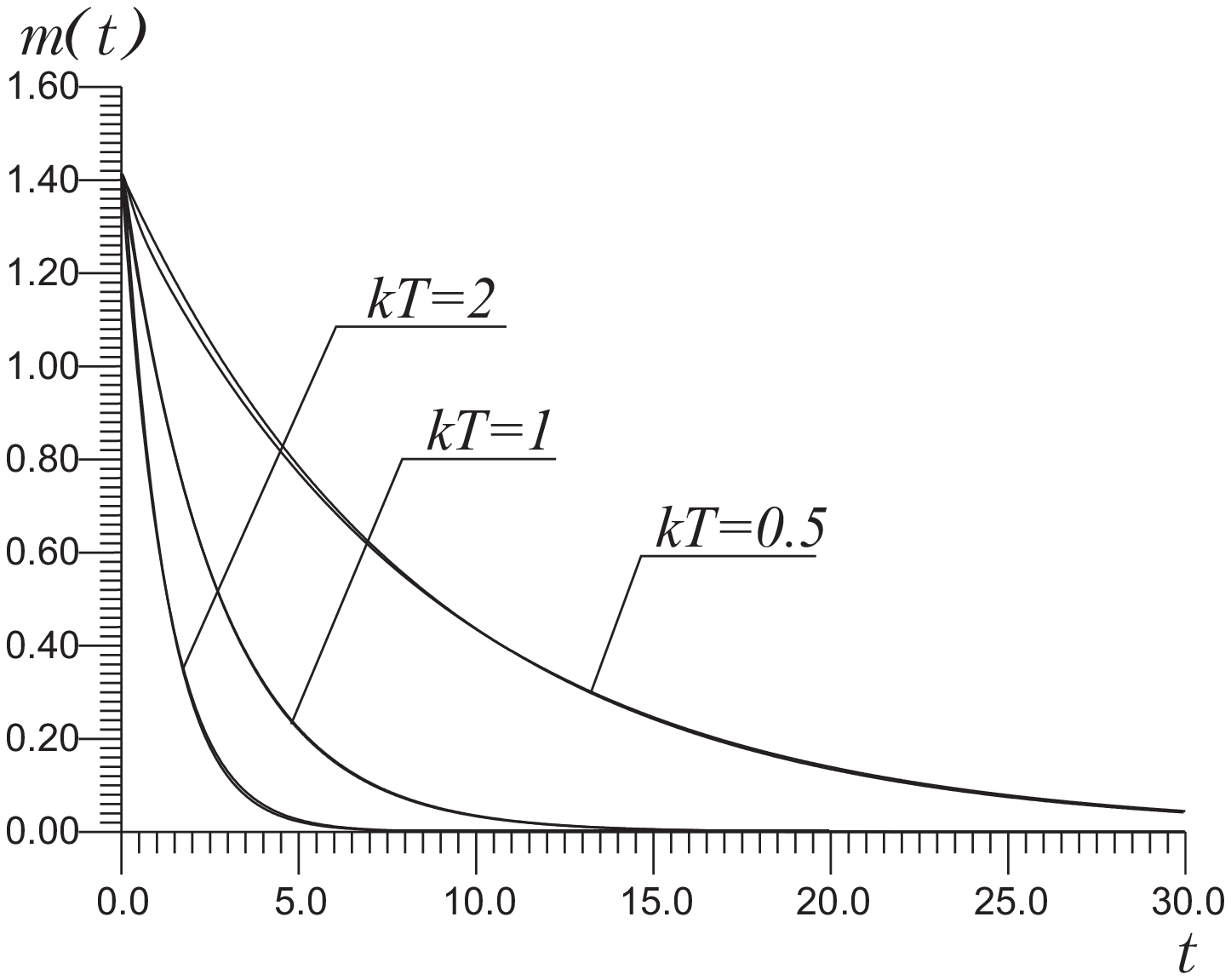}}
\vspace{-55pt}
\caption[b]{Evolution of the mean coordinate in the potential
${\mit\Phi(x)}=x^4/4-x^2$
for different values of noise intensity with initial
distribution located in a potential minimum. }
\end{figure}

\newpage

\begin{figure}[th]
\centerline{
\epsfxsize=12cm
\epsffile{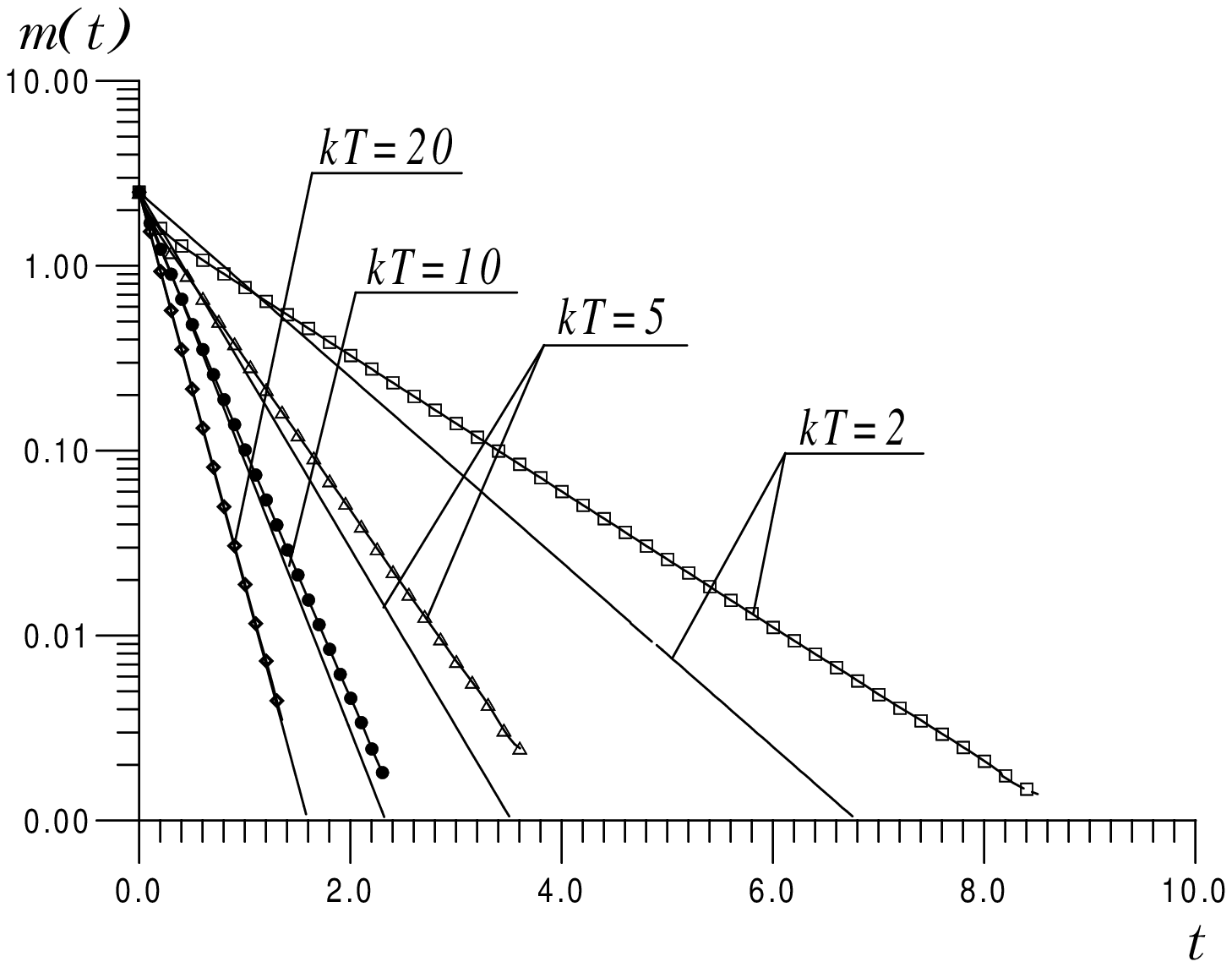}}
\vspace{-55pt}
\caption[b]{Evolution of the mean coordinate in the potential
${\mit\Phi(x)}=x^4/4-x^2$
for different values of noise intensity with initial
distribution located far from a potential minimum.
}
\end{figure}

}
\end{document}